\documentclass{mn2e}
\usepackage{amssymb}

\usepackage{times}
\usepackage{epsfig}

\def\be{\begin{equation}}
\def\ee{\end{equation}}
\def\bc{\begin{center}}
\def\ec{\end{center}}
\def\beq{\begin{eqnarray}}
\def\eeq{\end{eqnarray}}

\def\msun{{\rm M_\odot}}
\def\xte{{\it RXTE}}
\def\rinf{r_{\rm spot, \infty}}
\def\rspotc{r_{\rm spot, c}}
\def\Tinf{T_{\infty}}
\def\d{{\rm d}}
\def\rg{r_g}
\def\betaeq{\beta_{\rm eq}}
\def\Dop{\delta}
\def\taut{\tau_{\rm es}}
\def\source{SAX J1808.4$-$3658}

\title[X-rays from Accreting Millisecond Pulsar SAX J1808.4$-$3658]
{On the Nature of the X-ray Emission from the 
Accreting Millisecond Pulsar SAX~J1808.4$-$3658}

\author[Juri Poutanen and Marek Gierli\'nski]
{\parbox[]{6.8in} {Juri~Poutanen$^{1}$\thanks{E-mail:
juri.poutanen@oulu.fi (JP); Marek.Gierlinski@durham.ac.uk (MG)} 
\thanks{Corresponding Fellow, NORDITA, Copenhagen}
and  Marek Gierli\'nski$^{2,3}$\footnotemark[1]}\\
$^{1}$ Astronomy Division, P.O. Box 3000, FIN-90014 University of Oulu,
Finland  \\
$^{2}$ Department of Physics, University of Durham, South Road, 
Durham DH1 3LE, UK \\
$^{3}$ Astronomical Observatory, Jagiellonian University, Orla 171, 
30-244 Krak\'ow, Poland}

\begin{document}

\date{Accepted, Received}


\maketitle
\label{firstpage}


\begin{abstract}
The pulse profiles of the accreting X-ray millisecond pulsar \source\ at
different  energies are studied.  The two main emission  component,  the
black body and the Comptonized  tail that are clearly  identified in the
time-averaged spectrum, show strong variability with the first component
lagging the second one.  The observed  variability  can be  explained if
the  emission is produced  by  Comptonization  in a hot slab  (radiative
shock) of  Thomson  optical  depth  $\sim$  0.3--1 at the  neutron  star
surface.  The  emission  patterns of the black body and the  Comptonized
radiation are different:  a ``knife"-- and a ``fan"--like, respectively.
We construct a detailed model of the X-ray production accounting for the
Doppler  boosting,  relativistic   aberration  and  gravitational  light
bending in the  Schwarzschild  spacetime. We  present also accurate  analytical
formulae for  computations  of the light  curves from  rapidly  rotating
neutron stars using formalism recently developed by Beloborodov  (2002).
Our model reproduces well the
pulse profiles at different energies simultaneously, corresponding phase
lags, as well as the  time-averaged  spectrum. We constrain the compact
star mass to be bounded between $1.2$ and $1.6\msun$.
By fitting the  observed profiles, we determine
the radius of the compact  object to be $R\sim11$
km if $M=1.6\msun$,  while for $M=1.2\msun$ the best-fitting
radius is $\sim6.5$ km, indicating  that the compact  object in \source\
can be a strange  star.  We obtain a lower limit on the inclination of the
system of $65\degr$.
\end{abstract}

\begin{keywords}
accretion, accretion discs --  methods: data analysis -- 
pulsars: individual (SAX J1808.4$-$3658) -- X-rays: binaries
\end{keywords}

\section{Introduction}

Discovery of coherent  oscillations with frequencies in the 300--600
Hz range during X-ray  bursts from a number of low-mass  X-ray  binaries
(see Strohmayer \& Bildsten 2003 for a review) has triggered the efforts
to use the  information on the amplitude of  variability  and the folded
pulse shape to put  constraints on the  compactness  of the neutron star
and thus its equation of state as well as the emission  pattern from the
neutron  star  surface  (e.g.  Miller \& Lamb  1998).  However, a rather
limited  photon  statistics  does not  allow  to  reach a high  accuracy
(Strohmayer  et al.  1997; Nath,  Strohmayer \& Swank 2002; Muno, \"Ozel
\& Chakrabarty 2002).

The  millisecond coherent pulsations discovered in the persistent 
emission of the four sources: \source\ (with the period $P=2.5$ ms; 
Wijnands \& van der Klis 1998a), 
XTE J1751$-$305 ($P=2.3$ ms; Markwardt et al. 2002), 
XTE J0929$-$314 ($P=5.4$ ms; Galloway et al. 2002) and 
XTE J1807$-$294 ($P=5.25$ ms; Markwardt, Smith \& Swank 2003)
allow one to increase the statistics by folding the profile 
over a longer observational period
(days rather than seconds in the X-ray burst oscillations). 
Thus, for example, \source\ showed coherent pulsations with rms
amplitude of 5--7 per cent during its April 1998 outburst (Cui et al. 1998)
with almost constant shape of its energy spectrum 
(Gilfanov et al. 1998) and very similar pulse profiles. 
The pulse profiles of \source\ show strong energy dependence with soft 
photons lagging behind the hard ones (soft phase lags, see Cui et al. 1998). 
By analysing the  phase-resolved spectra,
Gierli\'nski, Done \& Barret (2002) showed that 
this results from the fact that 
the two main spectral components,
a soft black body and a hard Comptonized,  vary out of phase
with the first lagging the last one.  
While  the variations of the black body flux can be described by a 
single sine wave, the pulse of the hard Comptonized component is
strongly skewed. Gierli\'nski et al. (2002) suggested that 
the Doppler boosting can play a role in changing the shape of the 
profile.

The radiation pattern of standard  accreting X-ray pulsars is influenced
by strong  magnetic  field  $B\sim  10^{12}$ G.  On the other  hand, the
magnetic  field of the accreting  millisecond  pulsars in low mass X-ray
binaries is much weaker,  $B=10^8$--$10^9$  G (see e.g.  Wijnands \& van
der Klis  1998a;  Psaltis  \&  Chakrabarty  1999),  and does not  affect
significantly  properties of the emitted  radiation.  Thus these sources
can also serve as laboratories for studying  radiative  processes at the
surface of weakly-magnetized neutron stars.

In this paper, we construct a detailed model for the X-ray emission from
the  surface  of  a  rapidly   rotating   neutron  star  accounting  for
relativistic  effects.  We compare  the model with the data on  \source\
and put  constraints on the  inclination  of the system, the position of
the emitting  region  relative to the  rotational  pole and its emission
pattern as well as the stellar  radius.  The data used for the  analysis
are   described  in   \S~\ref{sec:data}.  The  model  and 
useful analytical formulae for light curve calculations are  presented  in
\S~\ref{sec:model}. The main results are given   in   \S~\ref{sec:results}.
The discussion   and the summary are given in
\S~\ref{sec:disc} and \S~\ref{sec:summary}, respectively.

\section{Data} 
\label{sec:data}

We study the data obtained by the {\it Rossi X-ray Timing Explorer}
(\xte) during the April 1998 outburst of \source, using {\sc ftools}
5.2. To improve the statistics we average the data between 
1998 April 11 and 29 (resulting in 118 ks of data).
For spectral fitting we extract the Proportional Counter
Array (PCA) spectra from all five detectors, top layer only, and use
the data in 3--20 keV energy band. We also extract spectra from
High-Energy X-ray Timing Experiment (HEXTE) in the 20--150 keV band, from
clusters 0 and 1. We constructed the energy-dependent pulse profiles
from the PCA data (all detectors, all layers) by following the
procedure described in Cui et al. (1998) and Gierli\'nski et al.
(2002) correcting the photon arrival times for orbital motions of the
pulsar and the spacecraft. 
For pulse profile fitting we created two
folded light curves in 16 phase bins, for energies 3--4 keV and 12--18
keV. The statistical uncertainties are about the same as 
uncertainties in the background, resulting in errors of 0.2 per cent and 0.3 per cent, 
respectively,  in the count rate in the two considered energy bands.

\section{Model}
\label{sec:model}

\subsection{Calculational method}

Accreting matter following the magnetic field lines close to the neutron
star is  stopped  in the  very  vicinity  of the  surface  by  radiation
producing  radiation  dominated  shock (Basko \& Sunyaev 1976;
Lyubarskii \& Sunyaev 1982).  For the
source luminosity of a few percent of the Eddington luminosity (Gilfanov
et al.  1998;  Gierli\'nski et al.  2002), the vertical (i.e.  along the
radial  direction)  extension  of the shocked  plasma is smaller  than a
characteristic  horizontal  size and certainly  smaller than the stellar
radius.  The  hard  X-rays  produced  in the  shock  can  irradiate  the
surrounding  stellar surface, so that the black body emission region can
cover a somewhat larger area.  For our  calculations  we assume that all
photons  originate  at the  stellar  surface  (i.e.  the  height  of the
emitting region is set to zero).

We  assume a  circular  spot and  consider  two  extreme  cases  for the
relative positions of the black body and the Comptonizing  regions:  (1)
a homogeneous slab, i.e.  the hot Comptonizing  layer covering the whole
black body spot, (2) a point--like  Comptonizing region in the centre of
a  black body region.

In order to compute the light curve as observed  by a distant  observer,
we first specify the radiation  spectrum and angular  dependence  in the
frame co-rotating with the star.  We then make Lorentz  transformation to
obtain the radiation  intensity in the  non-rotating  frame close to the
stellar  surface.  At the final step, we follow photon  trajectories  in
the Schwarzschild spacetime to infinity.

\begin{figure}
\centerline{\epsfig{file=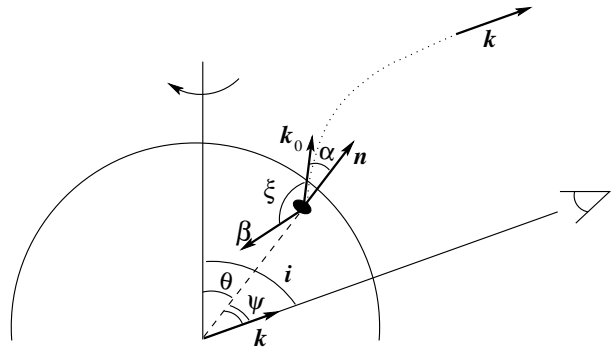,width=8.0cm}}
\caption{ The geometry of the problem. Dotted curve shows the 
photon trajectory. 
\label{fig:geom}}
\end{figure}

Let us now  consider  a  star  with  an  azimuthally  symmetric  surface
radiation  intensity  $I'(E',\alpha')$  that can vary over the  surface,
where  $\alpha'$  is the angle  between an emitted  photon and the local
normal to the stellar surface.  Let $\d S'$ be a surface element (spot) at
colatitude   $\theta$   (see
Fig.~\ref{fig:geom} for the geometry).  The primed quantities correspond
to the frame  co-rotating  with the spot.  Let $\bmath{k}$  and  $\bmath{n}$ be unit
vectors pointing from the star centre towards the observer and the spot,
respectively,  and  $i$  be  the  inclination  of  the  spin  axis.  The
inclination of the spot varies periodically
\be \label{eq:psi}
\cos\psi=\bmath{k}\cdot\bmath{n}=\cos i\ \cos\theta+\sin i\ \sin \theta\ \cos\phi,
\ee
where the phase $\phi=2\pi\nu t$, with $\nu$ being the pulsar frequency and
$t=0$ is chosen when the spot is closest to the observer. 

We compute the original direction of the photon $\bmath{k}_0$ near the 
stellar surface (which is transformed to $\bmath{k}$ at large distance 
from the star) assuming Schwarzschild geometry where   
photon orbits are planar: 
\be\label{eq:k0} 
\bmath{k}_0=[ \sin\alpha\ \bmath{k} +\sin(\psi-\alpha)\ \bmath{n}]/\sin\psi,  
\ee
where $\alpha$ is the angle between $\bmath{k}_0$ and $\bmath{n}$, 
i.e. $\cos\alpha=\bmath{k}_0\cdot\bmath{n}$.  

The relation between $\alpha$ and $\psi$ (i.e. light bending) 
can be obtained by standard techniques  (Pechenick,  Ftaclas \& Cohen 1983):
\be \label{eq:bend} 
\psi=\int_R^{\infty} \frac{\d r}{r^2} \left[ \frac{1}{b^2} - 
\frac{1}{r^2}\left( 1- \frac{\rg}{r}\right)\right]^{-1/2} ,  
\ee 
where the impact parameter
\be \label{eq:impact} 
b=\frac{R}{\sqrt{1-\rg/R}} \sin\alpha ,   
\ee
$\rg=2GM/c^2$ is the Schwarzschild radius, 
$M$ is the  mass and $R$ is the radius of the compact star. 

\begin{figure*}
\centerline{\epsfig{file=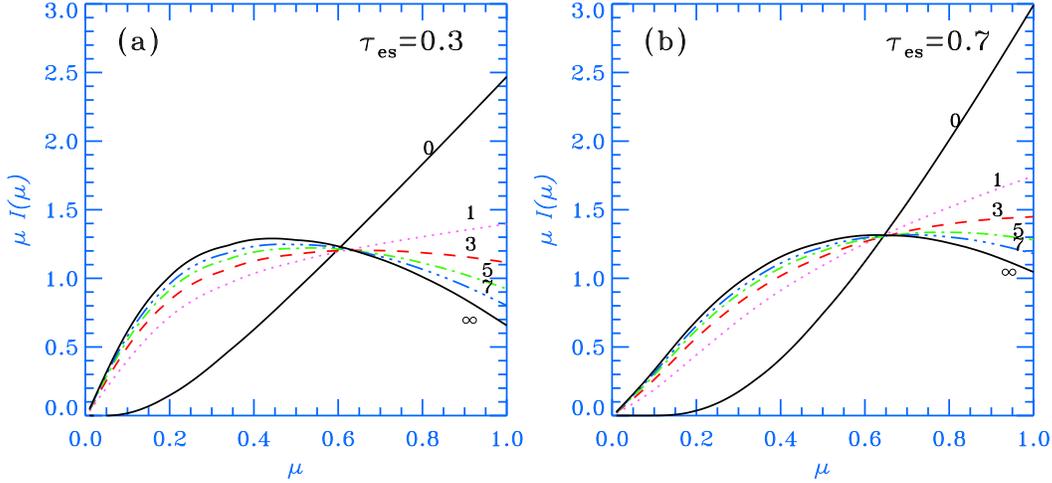,width=14.0cm}}
\caption{
Angular distribution of the radiation flux (normalized as 
$\int \mu I(\mu) d\mu=1$)
escaping from the top of an electron scattering
slab of Thomson optical depth $\taut=0.3$ ({\it left}) and $\taut=0.7$ 
({\it right}).
Incident radiation from the bottom is a black body (intensity independent of
the zenith angle $\arccos \mu$).
Computations follow procedure described in Sunyaev \& Titarchuk (1985).
Different scattering orders are shown and marked by numbers.
In the Comptonization process, scatterings also shift photons along the
energy axis so that at a given energy one scattering order dominates
(see Sunyaev \& Titarchuk 1985; Poutanen \& Svensson 1996). 
\label{fig:ray}}
\end{figure*}

The observed flux at energy $E$ is $\d F_E=I(E) \d\Omega$, where 
$I(E)$ is the radiation intensity at the infinity and 
$\d\Omega$ is the solid angle
occupied by $\d S'$ on the observer's sky. 
The solid angle can be expressed through the impact parameter 
\be \label{eq:omegab} 
\d\Omega=b\ \d b\ \d\varphi/D^2, 
\ee 
where $D$ is the distance to the source and 
$\varphi$ is the azimuthal angle corresponding to rotation around 
$\bmath{k}$. The impact parameter $b$ depends on $\psi$ only, but not
on $\varphi$.  

The apparent area of the spot as measured by photon beams
in the non-rotating frame near the stellar surface is $\d S=\Dop\ \d S'$
(see Terrell 1959; Lightman et al. 1975; Ghisellini 1999) 
and the relation between $\alpha$ and $\alpha'$ is described 
by the relativistic aberration formula (for motions parallel to the 
spot surface) 
\be\label{eq:aberr}
\cos\alpha'=\Dop\ \cos\alpha , 
\ee
where $\Dop$ is the Doppler factor.
Thus the spot area projected on to the plane perpendicular 
to the photon propagation direction, i.e. a photon beam cross-section,
is Lorentz invariant (see e.g.  Lightman et al. 1975; 
Lind \& Blandford 1985): 
\be \label{eq:Salpha}
\d S\ \cos\alpha = \d S'\ \cos\alpha' . 
\ee
Representing $\d S=R^2\ \d\cos\psi\ \d\varphi$ 
and using equations~(\ref{eq:impact}) and (\ref{eq:Salpha}) we get 
from (\ref{eq:omegab})
\be\label{eq:omega}
\d\Omega=\frac{\d S' \cos \alpha'}{D^2}\ \frac{1}{1-\rg/R}\ \frac{\d\cos\alpha}{\d\cos\psi}.
\ee
The Doppler factor can be expressed as follows 
\be \label{eq:dop}
\Dop=1/[\gamma(1-\beta\cos\xi)] , 
\ee 
where $\gamma=(1-\beta^2)^{-1/2}$ and 
$\beta=v/c$ is the spot velocity  as measured in the non-rotating
frame, 
\be\label{eq:beta}
\beta=\frac{2\pi R}{c} \frac{\nu}{\sqrt{1-\rg/R}} \sin\theta
=\betaeq \sin\theta .
\ee
Here $\betaeq$ is the velocity at the equator
and $\xi$ is the angle  between the spot velocity and $\bmath{k}_0$.
The pulsar frequency is corrected here for the redshift $\sqrt{1-\rg/R}$.
Using equation~(\ref{eq:k0}) it is easy to show that
\be \label{eq:cosxi}
\cos\xi=\frac{\bbeta}{\beta} \cdot \bmath{k}_0
=\frac{\sin\alpha}{\sin\psi} \frac{\bbeta}{\beta} \cdot \bmath{k}=
- \frac{\sin\alpha}{\sin\psi}\sin i\ \sin\phi\  .
\ee

In the case of a weak field and negligible bending 
the solid angle occupied by the spot is 
$\d\Omega=\d S' \cos \alpha'/D^2$. 
In reality, this formula can be applied even when bending is 
significant since the relation between $\cos\alpha$ and $\cos\psi$ is close
to linear (Beloborodov 2002), 
\be\label{eq:cosbend}  
\cos\alpha\approx \rg/R + (1-\rg/R) \cos\psi , 
\ee 
for a star with $R>2\rg$, so that $\d\cos\alpha/\d\cos\psi\approx 1-\rg/R$.  

The    radiation    intensity    observed    at    the    infinity    is
$I(E)=(1-\rg/R)^{3/2}  I_0(E_0,\alpha)$, where $E/E_0=\sqrt{1-\rg/R}$ is
the redshift  (Misner,  Thorn \& Wheeler  1973) and the energy  $E_0$ is
measured in the  non-rotating  frame near the stellar  surface.  (In the
actual calculations, one can neglect the redshift factor in all formulae
for photon energies, since it is the same for the photons originating at
the stellar  surface.)  Now we can make  Lorentz  transformation  to the
spot rest frame where we know the angular and energy distribution of the
radiation  field  $I'(E',\alpha')$.  The  intensities  are  related  via
$I_0(E_0,\alpha)=\Dop^3   I'(E',\alpha')$ with the   Doppler   shift  
given by $E_0=\Dop\ E'$.

Combining equations above we get the expression for the
observed flux:
\be  \label{eq:fluxspot}
\d F_E= \frac{\sqrt{1-\rg/R}}{D^2} \ \Dop^3  I'\left( E',\alpha'\right)
\frac{\d\cos\alpha}{\d\cos\psi}  \  \d S' \cos \alpha'  ,
\ee
with the visibility condition  $\cos \alpha'>0$.
We  account  thus here for the  special  relativistic  effects  (Doppler
boosting, relativistic aberration), the gravitational redshift and light
bending in  Schwarzschild  geometry.  In order to evaluate the flux at a
given  phase   $\phi$,  one  first   computes   $\psi$  using   equation
(\ref{eq:psi}),  then  inverting   (\ref{eq:bend})   gets  $\alpha$  and
substitutes   the   results   into   equations    (\ref{eq:cosxi})   and
(\ref{eq:dop})  to obtain  the  Doppler  factor  which is used  later to
compute $\alpha'$ from equation  (\ref{eq:aberr}).  The light curve from
the antipodal  spot can be easily  obtained by replacing in all formulae
$\theta\rightarrow \pi-\theta$ and $\phi \rightarrow \pi+\phi$.  We note
here that the formalism  outlined above  neglects the time delays due to
different  photon paths (see Pechenick et al.  1983) which can be easily
incorporated.  However,  they are  negligible  comparing  to the  pulsar
period above a few ms.

For a spot of the  finite  size, the observed flux can be  obtained  by
integrating over the (visible) spot area.  On the other hand, if we know
from the observations  the black body flux $F_{\rm bb}$, its temperature
$\Tinf$ and the distance to the source $D$, one can estimate the size of
the (black body) emitting region.  The  ``observed''  radius of the spot
$\rinf=(F_{\rm  bb} D^2 /  \sigma_{\rm  SB} \Tinf^4  )^{1/2}$  should be
corrected  for the effects of gravitation and orientation  (using  a
procedure  described in  \S~\ref{sec:appr}, see also Psaltis, \"Ozel \&
DeDeo 2000) to obtain the actual spot area.

\begin{figure}
\centerline{\epsfig{file=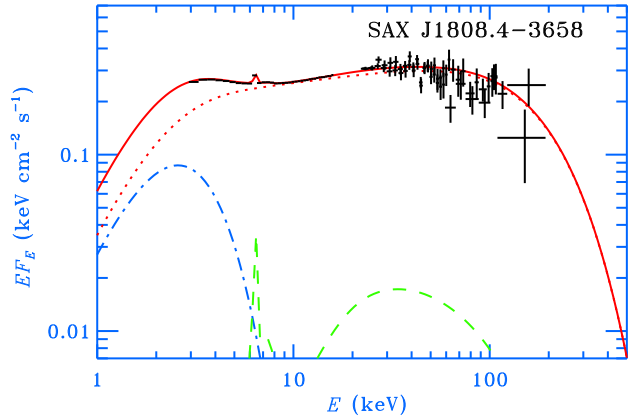,width=8.0cm}}
\caption{
Time-averaged spectrum of \source\ as observed
by \xte. The model spectrum  (solid curve)
consists of a black body (dashes/dots),
thermal Comptonization (dots),
and Compton reflection with the iron line (dashes).
Thermal Comptonization component (model {\sc thcomp} in {\sc xspec};
Zdziarski, Johnson \& Magdziarz 1996) has a best-fitting photon 
spectral slope $\Gamma=1.86$ 
(the electron temperature was fixed at 90 keV) and the Compton reflection 
amplitude $\Omega/(2\pi)=0.09$. 
The black body modeled by {\sc bbodyrad} from {\sc xspec} has the best-fitting
temperature $kT_{\rm bb}=0.66$ keV and the radius of the circular 
emitting region $\rinf=2.4$ km (for the source distance $D=2.5$ kpc).  
\label{fig:speave}}
\end{figure}

In our  calculations  of the model light curves for  \source\ we use the
distance  $D=2.5$  kpc (in't Zand et al.  2001),  $\Tinf=0.66$  keV, and
$\rinf=2.4$  km  (see  Fig.~\ref{fig:speave}  and  Gierli\'nski  et  al.
2002).  Since the observed profiles from \source\ have almost sinusoidal
shapes, they cannot be produced by two antipodal  spots.  Therefore,  we
compute the light curves from the primary spot (closest to the observer)
only.  We  also neglect  multiple  images
since  neutron or strange stars have sizes larger than $1.5 \rg$
(see e.g. Lattimer \& Prakash 2001).

\subsection{Incident spectrum and its angular distribution}

In order to compute a light curve $F_E(\phi)$ at a given energy $E$, one
needs to specify the intrinsic  radiation pattern in the spot rest frame
$I'(E',\alpha')$.  In a slab geometry of the emitting  region, the black
body  photons  come from the bottom of the hot  Comptonizing  slab.  The
black body  intensity  transmitted  through the slab of Thomson  optical
thickness  $\tau=n_e\sigma_T h$ is  $I^{\rm  bb}(\mu)\propto  \exp(-\tau/\mu)$,  where
$\arccos\mu$  is the zenith (polar) angle measured from the slab normal,
$n_e$ is the electron concentration, $\sigma_T$ is Thomson cross-section,
and $h$ is the height of the slab.
The radiation  flux from a unit area is then  strongly  peaked along the
normal direction (a ``knife''--like  emission pattern, see curves marked
by 0 at  Fig.~\ref{fig:ray}).  The  scattered  radiation  is expected to
have a  very  different  angular  distribution,  which  depends  on  the
scattering  order (and thus on photon  energy).  The radiation flux does
not peak  anymore in the normal  direction,  but  instead  peaks at some
intermediate  angle (a ``fan''--like  emission pattern, see solid curves
marked by $\infty$ in Fig.~\ref{fig:ray}).  It approaches the asymptotic
distribution  $I_{\infty}(\tau,\mu)$  in a few  scatterings  (the  exact
number is a function of the optical depth). We compute the angular
distribution of radiation following procedure described in Sunyaev \&
Titarchuk (1985).

The intensity of the black body radiation in the frame
co-moving with the spot can be represented in the following way:
\be\label{eq:bb} 
I^{\rm bb}_{E'}(\alpha')\propto \exp(-\tau/\cos\alpha')
f_{\rm bb}(E'\sqrt{1-\rg/R})  , 
\ee
where  $f_{\rm bb}(E)$ is the observed time-averaged black body flux
(i.e. the best-fitting {\sc bbodyrad} model spectrum, see Fig.~\ref{fig:speave}).
We assume that 
the energy and angular dependences of the high energy (scattered many times) 
radiation can be separated as 
\be\label{eq:sc}
I^{\rm sc}_{E'}(\alpha') \propto I_{\infty}(\tau,\alpha')f_{\rm sc}(E'\sqrt{1-\rg/R}) .
\ee 
Here $f_{\rm sc}(E)$ is the observed, time-averaged energy dependence of the
Comptonized flux 
(i.e. the best-fitting {\sc thcomp} model spectrum, see Fig.~\ref{fig:speave}). 

Since the main  spectral  components  of the  time-averaged  spectrum of
\source\ are a power-law--like  Comptonized component (with weak Compton
reflection) extending at least to 200 keV and a black body contributing
about   30 per cent   to   the   flux   in   the   3--5   keV   region,
(see Fig.~\ref{fig:speave}  and  Gilfanov et al.  1998;  Gierli\'nski  et al.
2002),  we  compute  the  total  observed  flux as a  function  of phase
$F_E(\phi)$  at every energy $E$ summing up the  contributions  from the
black   body   and  the   Comptonized   components.  We   compute   them
independently  and renormalise so that their  phase-averaged  values are
equal  to the  observed  fluxes  $f_{\rm  bb}(E),  f_{\rm  sc}(E)$  (see
Fig.~\ref{fig:speave}).

If the black  body  spot is not  covered  completely  by a  Comptonizing
layer, some black body photons do not enter the Comptonizing  region and
escape to the observer directly.  In that case, the ``effective'' $\tau$
that   describes  the  mean  (i.e.  averaged   over  the  spot)  angular
dependence  of the black body  radiation  can be smaller than the actual
optical depth in the  Comptonizing  region  $\taut$ that  describes  the
angular  dependence  of  the  Comptonized  radiation.  Since  the  exact
geometry  is  unknown,  we use $\tau$  and  $\taut$  as two  independent
parameters for the light curve computations.  The emission pattern for 
the black body radiation is given
by equation~(\ref{eq:bb}) while the Comptonized radiation is 
described by equation~(\ref{eq:sc}) where instead of $\tau$ we use 
$\taut$ as a parameter for the data  fitting. 
These patterns are referred to as  model  1  in   \S~\ref{sec:results}   and
Table~\ref{table:fit}.

In order to check the robustness of the results and their  dependence on
the assumed  angular  distribution  of  radiation,  we consider a second
model where instead of $I_{\infty}(\taut,\mu)$  we use a linear function
$1+a\mu$  with $a$ as a free  parameter  (limited  by $-1$ from below to
ensure  positiveness  of  the  function).  The  corresponding  fits  are
referred below as model 2.

Thus, the model  parameters  are the  neutron  star mass $M$, its radius
$R$,  rotational  frequency  $\nu$ ($=401$ Hz for \source),  inclination
$i$, colatitude of the spot centre $\theta$,
the optical  depth $\tau$ and a  parameter  determining  the
angular distribution of the Comptonized radiation $\taut$ or $a$ (plus a
free phase factor).

\subsection{Analytical light curves}
\label{sec:appr}

Using the formalism developed by Beloborodov (2002) we  derive here 
simple analytical formulae for the light curves and the oscillation 
amplitudes that could be useful for understanding the 
main physical effects. Let us first consider  a
small spot and make  approximation  to the light bending  formulae.  The
bolometric  flux can be obtained  from  equation~(\ref{eq:fluxspot})  by
integrating  over $E$ and using the cosine  relation  (\ref{eq:cosbend})
which describes well the light bending near a star with $R\gtrsim  2\rg$:
\beq  \label{eq:dF1}
\d F&=& (1-\rg/R)^2 \Dop^4 I'(\alpha') \Dop\cos\alpha\   \d S'/D^2 \\
&=& \left( 1-\frac{\rg}{R} \right)^2
\Dop^5 \left[ \frac{\rg}{R}+ \left(1-\frac{\rg}{R}
\right)  \cos\psi  \right] I'(\alpha') \frac{\d S'}{D^2} , \nonumber
\eeq
with the visibility condition  $\cos\psi>\mu_v\equiv -\rg/(R-\rg)$.  (For
the  antipodal  spot,  one  should  substitute   $\cos\psi   \rightarrow
-\cos\psi$ and $\phi\rightarrow -\phi$ when computing the Doppler factor.)

\begin{figure}
\centerline{\epsfig{file=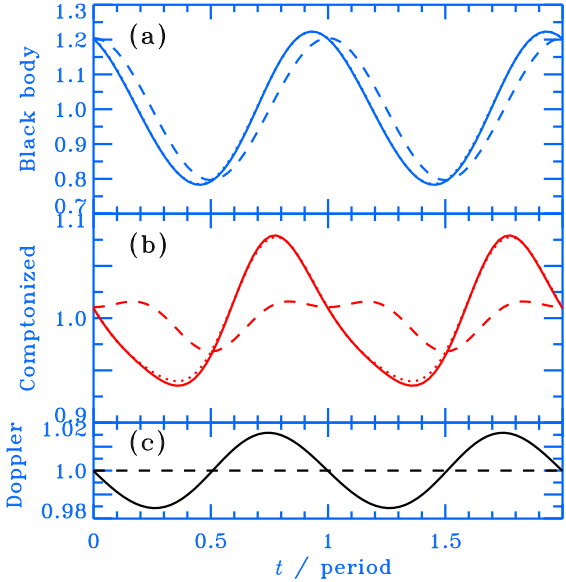,width=7.5cm}}
\caption{
The bolometric black body and Comptonized fluxes, and the Doppler
factor $\Dop$ as functions of phase.
Dashed curves correspond to a slowly rotating neutron star.
Solid curves give the profiles modified by the Doppler boosting and aberration
for a neutron star of rotational frequency 401 Hz.
(Other parameters are given in Fig.~\ref{fig:all}.)
Light bending in the Schwarzschild geometry is accounted for.
Dotted curves (coinciding completely with the solid one
at the upper panel) show the results of multiplication of the dashed curves by
$\Dop^5$ and $\Dop^4$ for the black body and the Comptonized emission,
respectively (see eqs [\ref{eq:dF1}] and [\ref{eq:Fsc}]).
\label{fig:relateff}}
\end{figure}

It is clear from equation  (\ref{eq:dF1})  that the (bolometric) 
black body flux
observed  from a spot at a rapidly  spinning  star is a factor  $\Dop^5$
times the flux from a slowly rotating star
(see Fig.~\ref{fig:relateff}a).  Two powers of $\Dop$ come
from the solid angle transformation, one from the energy, one from the
arrival time contraction, and the fifth from the change in the projected  area due
to  aberration.  The  scattered  radiation,  on the  other  hand, has an
angular  distribution  which  differs from that of the black body.  In
the      optically      thin     case,     it     is     closer     to
$I'(\alpha')\cos\alpha'=f'\approx  const$ (see  Fig.~\ref{fig:ray}). 
Equation (\ref{eq:dF1}) for the observed flux can be rewritten then as
\be \label{eq:Fsc}
\d F_{\rm sc}=(1-\rg/R)^2 \Dop^4 f'\d S'/D^2.
\ee
One sees that the flux is modified by approximately  factor  $\Dop^4$
(see Fig.~\ref{fig:relateff}b).  The angular
distribution of the hard radiation (defined in our model by parameters
$\taut$ or $a$) is  reflected  in the shape of the light curve at high
photon  energies, where  Comptonizing  emission  dominates.  Since the
Comptonization   spectrum  is  close  to  a   power-law   $I(E)\propto
E^{-(\Gamma-1)}$,  with the  same  approximations  as  above,  one can
easily get the monochromatic flux at a given energy:
\be 
\d F^{\rm sc}_{E}=(1-\rg/R)^{3/2} \Dop^{2+\Gamma} f'(E_0) \d S'/D^2 ,
\ee
which varies in a way similar to the bolometric flux if $\Gamma\approx 2$.

Neglecting the Doppler effect and  using equation~(\ref{eq:dF1})
one can relate the variability amplitude 
of the black body flux
to the angles $i$ and $\theta$ and the neutron star radius
(Beloborodov 2002):
\be \label{eq:ampl}
A \equiv (F_{\max}-F_{\min})/(F_{\max}+F_{\min})= U/Q ,
\ee
where we defined
\beq \label{eq:uq}
U&=&(1-\rg/R) \sin i\ \sin \theta , \nonumber \\
Q&=&\rg/R+(1-\rg/R) \cos i \ \cos\theta .
\eeq
This peak-to-peak amplitude  is $\sqrt{2}$ larger than
sometimes quoted rms amplitude if the light curve is a pure sine wave.

For a circular spot of an angular radius   $\rho$
emitting as a black  body,
the observed flux can be obtained from  equation  (\ref{eq:dF1}):
\be\label{eq:fluxrho}
F_{\rho}= \left( 1-\frac{\rg}{R}\right)^2 \frac{I_0}{D^2} \pi R^2  \sin^2\rho
\left( Q+ \frac{\rg}{R} \tan^2\frac{\rho}{2} + U \cos\phi \right) ,
\ee
if  the spot is  always   visible, i.e. $\cos(i+\theta+\rho)>\mu_v$.
Here  the intensity is related to the local temperature
$I_0=\sigma_{\rm SB}T_0^4/\pi$.
We can easily now obtain the oscillation amplitude:
\be \label{eq:amplspot}
\strut\displaystyle
A_\rho=\frac{U}{\strut\displaystyle Q+\frac{\rg}{R} \tan^2\frac{\rho}{2}}.
\ee
Equation (\ref{eq:fluxrho})  allows us to estimate the size of
the spot $R\rho$  from the equation:
\be
\rinf^2=R^2 \sin^2\rho \left( Q+ \frac{\rg}{R} \tan^2\frac{\rho}{2} \right) ,
\ee
if the  distance  $D$, the temperature
$\Tinf=T_0  (1-\rg/R)^{1/2}$ and the observed phase-averaged black body
flux $\overline{F_\rho}=\sigma_{\rm SB} \Tinf^4 \rinf^2/D^2$  (and thus
$\rinf$) are known from observations.
For small $\rho$, the spot radius is $R\rho=\rinf Q^{-1/2}$.

\section{Results}
\label{sec:results}

We fit the  observed  pulse  profiles  in the energy  bands 3--4 keV and
12--18 keV  simultaneously.  The second band is completely  dominated by
the Comptonized  radiation and thus gives direct  information  about its
angular  distribution.  In the lower energy band, the black body affects
the  observed  pulse  profile.  Since  the mass of the  compact  star in
\source\  is not known, we  consider  masses  between  $M=1.0\msun$  and
$1.6\msun$.  Masses in the range of $1.4$--$1.6\msun$ 
are expected  for neutron  stars in low mass
X-ray binaries that have accreted material from their binary companions,
lower masses are possible, for example, for strange  stars.  We restrict
the  inclination to $i<82\degr$ to be consistent  with the absence of the
X-ray  eclipses   (Bildsten  \&  Chakrabarty   2001).  The  best-fitting
parameters  are  presented  in  Table~\ref{table:fit}.  The two  sets of
errors    correspond   to  the 90 and   99 per cent    confidence    intervals
with $\Delta\chi^2=2.71$   and   $6.63$,   respectively.

\begin{table*} 
\caption{Best-fitting parameters}
\begin{tabular}{cllllccc}
\hline
$M$        & $R$  & $R/\rg$ & $i$    &  $\theta$ & $\tau$ & $\taut$ or  $a$  & $\chi^2/$dof \\
($\msun$)  & (km) &         & (deg)  &  (deg)    &        &      &              \\
\hline
\multicolumn{8}{c}{Model 1:  \hspace{1cm} 
Electron scattering slab, $I^{\rm sc}(\mu)\propto I_{\infty}(\taut,\mu)$} \\
\hline 
$1.0$ & $5.15^{+0.25}_{-0.15}\ ^{+0.45}_{-0.35}$ & $1.72^{+0.08}_{-0.05}\ ^{+0.15}_{-0.12}$ 
& $74^{+8}_{-20}\ ^{+8}_{-24}$ & $20^{+6}_{-2.5}\ ^{+10}_{-3.5}$ & 
        $0.08^{+0.07}_{-0.04}\ ^{+0.09}_{-0.05}$  & $0.84^{+0.22}_{-0.16}\ ^{+0.28}_{-0.19}$ &  $29.7/26$ \\ 
$1.2$ & $7.0^{+0.5}_{-0.4}\ ^{+1.1}_{-0.6}$ &  $1.94^{+0.14}_{-0.11}\ ^{+0.31}_{-0.17}$ 
 & $82^{+0}_{-5}\ ^{+0}_{-18}$ & $13^{+1.5}_{-1}\ ^{+3}_{-1.5}$ & 
        $0.06^{+0.04}_{-0.02}\ ^{+0.08}_{-0.03}$  & $0.59^{+0.09}_{-0.06}\ ^{+0.24}_{-0.10}$ &  $35.3/26$ \\  
$1.4$ & $8.8^{+2.9}_{-0.4}\ ^{+4.5}_{-0.7}$ & $2.1^{+0.7}_{-0.1}\ ^{+1.1}_{-0.17}$ 
& $82^{+0}_{-5}\ ^{+0}_{-8}$ & $11^{+0.5}_{-3}\ ^{+1}_{-4}$ &
        $0.08^{+0.03}_{-0.06}\ ^{+0.05}_{-0.07}$  & $0.52^{+0.05}_{-0.21}\ ^{+0.12}_{-0.22}$  & $47.3/26$ \\
$1.6$ & $13.4^{+0.8}_{-1.0}\ ^{+1.6}_{-1.7}$ & $2.8^{+0.17}_{-0.21}\ ^{+0.33}_{-0.35}$
& $82^{+0}_{-1}\ ^{+0}_{-2}$ & $7^{+1}_{-0.5}\ ^{+1.5}_{-1}$ &
        $0.05^{+0.02}_{-0.01}\ ^{+0.03}_{-0.01}$  & $0.33^{+0.04}_{-0.03}\ ^{+0.07}_{-0.03}$  &  $61.4/26$ \\
\hline
\hline
\multicolumn{8}{c}{Model 2:  \hspace{1cm} $I^{\rm sc}(\mu)\propto 1+a\mu$ } \\
\hline
$1.0$ & $5.0^{+0.2}_{-0.1}\ ^{+0.4}_{-0.2}$ & $1.67^{+0.07}_{-0.03}\ ^{+0.13}_{-0.07}$  
& $82^{+0}_{-25}\ ^{+0}_{-32}$ & $19^{+6}_{-1}\ ^{+11}_{-2}$ & 
        $0.09^{+0.07}_{-0.01}\ ^{+0.09}_{-0.01}$  & $-0.69^{+0.10}_{-0.02}\ ^{+0.12}_{-0.03}$ &  $28.1/26$ \\ 
$1.2$ & $6.4^{+0.3}_{-0.2}\ ^{+0.8}_{-0.3}$ & $1.78^{+0.08}_{-0.05}\ ^{+0.22}_{-0.08}$
& $82^{+0}_{-8}\ ^{+0}_{-17}$ & $14^{+1.5}_{-0.5}\ ^{+3}_{-1}$ & 
        $0.14^{+0.03}_{-0.02}\ ^{+0.06}_{-0.03}$  & $-0.74^{+0.06}_{-0.02}\ ^{+0.10}_{-0.04}$ &  $32.2/26$ \\  
$1.4$ & $8.3^{+0.6}_{-0.4}\ ^{+1.6}_{-0.7}$ & $1.97^{+0.15}_{-0.09}\ ^{+0.39}_{-0.16}$ 
& $82^{+0}_{-5}\ ^{+0}_{-13}$ & $11^{+1}_{-1}\ ^{+1.5}_{-1.5}$ &
        $0.14^{+0.03}_{-0.01}\ ^{+0.04}_{-0.02}$  & $-0.80^{+0.04}_{-0.04}\ ^{+0.08}_{-0.07}$  & $38.7/26$ \\
$1.6$ & $10.8^{+0.6}_{-0.8}\ ^{+2.6}_{-0.9}$ &   $2.25^{+0.13}_{-0.17}\ ^{+0.47}_{-0.27}$
& $82^{+0}_{-3}\ ^{+0}_{-7}$ & $8.5^{+1.0}_{-0.5}\ ^{+1.5}_{-1}$ &
        $0.15^{+0.02}_{-0.01}\ ^{+0.04}_{-0.01}$  & $-0.89^{+0.05}_{-0.01}\ ^{+0.08}_{-0.01}$  &  $44.8/26$ \\
\hline
\label{table:fit}
\end{tabular}
\end{table*}

We see from Table~\ref{table:fit} that the results obtained with the two
models 1 and 2 are very consistent  with each other.  For both models, a
large  inclination  is preferred and the lower limit is  $i>65\degr$  (at
99 per cent  confidence  level) for  $M\gtrsim 1.2\msun$.  The fits  become  much
worse at masses higher than  $1.6\msun$.  The estimated  stellar  radius
(even  in  $\rg$  units)   grows  with  the  mass  and  varies   between
$\sim1.7\rg$ and $\sim2.3\rg$  ($\sim2.8\rg$ for model 1).  Using simple
analytical  estimations we explain  the physical reasons behind this 
behaviour in  \S~\ref{sec:constr}. The radii are  consistent  with some of
the  neutron  star   equations  of  state  if   $M\gtrsim 1.4\msun$.
For a  smaller  assumed  mass of  $1.2\msun$,  the
radius is more consistent  with the equations of state for strange stars,
while for even smaller mass of $1\msun$,
the  obtained  stellar  radius of $R\sim 5$ km is too  small  even for a
strange  star (see \S~\ref{sec:massrad} and Fig.~\ref{fig:eos} for details).

The colatitude $\theta$ of the spot centre 
varies  approximately as $\propto  R^{-1}$.  A larger
$\theta$ is needed for  smaller  $R$ in order to keep the spot  velocity
(and thus the Doppler factor) at approximately constant level.

The estimated radius of the spot (not a fitting parameter) corrected for
orientation and gravitational  effects varies between 3.0 and 3.7 km for
$M=1.0\msun$  and $1.6\msun$,  respectively.  One can get some estimates
of the size of the  Comptonizing  region from a rather large  difference
between the  best-fitting  $\taut$ and $\tau$  obtained in model 1.  The
values of $\taut$ are consistent  with the optical depths  obtained from
the fits of the time-averaged spectrum with a Comptonization model for a
slab  geometry.  For example,  model {\sc compps}  (Poutanen \& Svensson
1996),\footnote{{\sc        compps}        is        available        at
ftp://ftp.astro.su.se/pub/juri/XSPEC/COMPPS}  gives  $\taut=0.5$ for the
electron  temperature  $kT_e=90$  keV  (which  is not  well  constrained
because of the low signal above 100 keV).  This leads us to a conclusion
that a large  fraction of the black body  photons  reaches the  observer
without  interaction  with the hot  Comptonizing  medium  (contradicting
assumed  homogeneous slab geometry).  In order to get the  ``effective''
$\tau\sim0.1$ (see eq.~[\ref{eq:bb}]),  about 40 per cent of the black body area
should be covered by the Comptonizing medium with $\taut\sim 0.5$, while
the  remaining  60 per cent  should  emit as a pure black  body  (i.e.  with no
$\exp(-\tau/\mu)$  reduction due to scattering).  Thus, we conclude that
the radius of the Comptonizing region $\rspotc$ is about 2 km.

\begin{figure}
\centerline{\epsfig{file=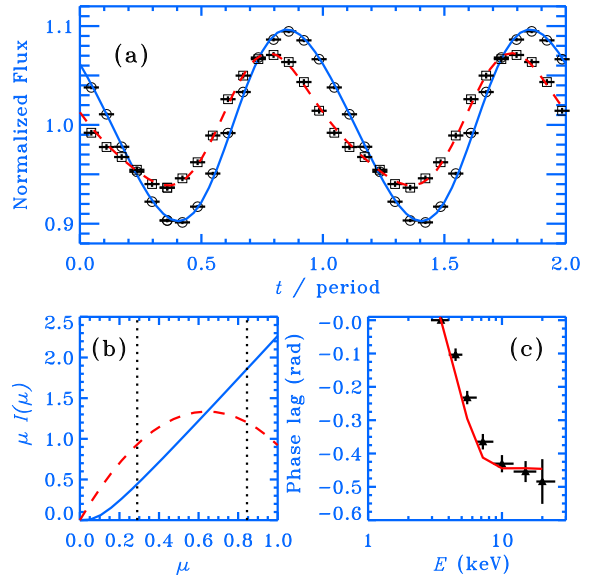,width=7.5cm}}
\caption{
(a) Observed normalized pulse profiles in the energy bands
3--4 keV (circles) and 12--18 keV (squares) and 
the best-fitting model 
light curves (solid and dashed curves) 
as a function of pulsar phase (two periods are plotted,
zero phase is chosen when a spot is closest to the observer).
A neutron star with the mass $M=1.4\msun$ and the radius
$R=2\rg=8.4$ km and inclination $i=80\degr$ are assumed.
The best-fitting parameters (model 2) are $\theta=11\degr$, $\tau=0.16$,
$a=-0.78$ and $\chi^2$/dof$=40.1/28$. 
(b) The angular distribution of the intrinsic black body (solid curve) and Comptonized
(dashed curve) fluxes $\mu I(\mu)$ (here $\mu=\cos\alpha'$)
in the co-rotating frame normalized as $\int \mu I(\mu) d \mu=1$.
Only the range of angles between the dotted lines is actually observed. 
(c) The observed (crosses) and the model (solid curve) phase lags
at the pulsar frequency relative to the 3--4 keV band.
\label{fig:all}}
\end{figure}

The angular  distributions  of the  scattered  radiation  (see  dashed
curve in  Fig.~\ref{fig:all}b)  determined by parameters  $\taut$ and
$a$ is also very similar for models 1 and 2.  Model 2 is somewhat less
restrictive and therefore gives  generally  better fits.  The required
emission  pattern is certainly very much different  from that given by
the Lambert law and more radiation is escaping at intermediate  angles
$\alpha'\sim50\degr$  than along the normal.  It is this pattern  which
is  responsible  for the origin of a strong first  harmonic  (i.e.  at
double spin  frequency)  in the pulsar  light  curve at high  energies
(modified  also by Doppler  effects, see  Fig.~\ref{fig:relateff}  and
\S~\ref{sec:appr}).

In order to check the dependence of the results on the assumed geometry,
we repeated the fitting  procedure  considering  an extreme  case with a
point-like  Comptonizing  source at the  centre of the black  body spot.
The  best-fitting  parameters  are similar to the  situation  where both
emitting  regions  coincide.  This is  expected  since  the  size of the
emission region should not play a large role since $r_{\rm spot}=R\rho<R$.
However, 
when estimating the size of the spot, we assumed 
the soft emission is a black body, i.e. the observed color temperature is 
equal to the effective temperature. For the hydrogen and helium atmospheres of 
weakly magnetized neutron stars one expects that 
the color correction is $f_{\rm col}\equiv T_{\rm col}/T_{\rm eff}=1.6$--$1.8$ 
(Zavlin, Pavlov \& Shibanov 1996).
The shift of the peak of the emission towards higher energies 
from the corresponding black body emission results from the fact that the 
outer layers of the atmosphere are cooler than the deeper layers. 
Since the bound-free and free-free cross-sections rapidly decrease with 
increasing photon energy, one sees deeper and hotter layers at high energies. 
However, when the accretion takes place and the radiative shock forms 
(its existence is supported by the presence of the hard X-ray tail), 
the soft radiation can be produced by reprocessing the hard photons 
from the shock. The temperature gradient of the atmosphere then 
can be opposite to that of standard neutron star atmospheres. Now 
the hot layers are at the top, softer photons are coming from the hotter 
region (since absorption cross-section is large for small energies), 
while hard photons are coming from the cooler 
region below and the color correction 
can be smaller than unity. The actual vertical temperature dependence 
is however unknown. 

To investigate the influence of different possible 
color corrections on our best-fitting parameters,
we consider two extreme cases of $f_{\rm col}=1.5$ and 
$0.7$ and fitted the data with model 2.  
The area of the ``black body'' spot now changes 
by a factor $f_{\rm col}^2$. 
For $f_{\rm col}=1.5$, the spot centre colatitude  $\theta$ increases by about 20 
per cent comparing to the case of no color correction. 
Since now the spot is larger, $\theta$ has to increase to keep
the same oscillation amplitude (see eq.~[\ref{eq:amplspot}]).
The best-fitting stellar radius increases significantly only 
for the smallest considered mass of $M=1\msun$ reaching $5.5$ km, 
for $M=1.2\msun$ the radius is $6.7$ km, 
while for higher  masses the change is negligible. 
In the case of $f_{\rm col}=0.7$ the spot becomes smaller and 
$\theta$ also slightly decreases. At $M=1\msun$ the best-fitting radius 
is now $4.8$ km, while all   
other parameters remain the same within the errors. 
There is no change in the best-fitting parameters for higher stellar masses. 
This results from the fact that  the light curve 
depends weakly on the spot size if it small comparing to the stellar radius 
(see eqs~[\ref{eq:fluxrho}, \ref{eq:amplspot}]).

One of the important assumptions that can affect our results is that the
two emission  components  are assumed to be  co-spatial  or are at least
co-centred.  In  our  interpretation  the  Comptonized   emission  peaks
earlier than the black body due to a different angular distribution more
affected by the Doppler  effect.  However, an additional  phase shift is
possible if the Comptonized emission region is physically shifted on the
stellar  surface.  We fitted the data with  model 2 and the phase  shift
$\Delta\phi$  of  the  Comptonized   emission  as  an  additional   free
parameter.  A major  effect  of  adding  a new  parameter  is  that  the
confidence   interval  for  the  stellar  radius  increases.  Thus,  for
example, for $M=1.4\msun$,  the best fit with  $\chi^2$/dof$=34/25$  was
obtained for  $\Delta\phi=0.1$  rad (a positive shift  corresponds to an
earlier  arrival  time)  and  other  parameters  very  similar  to those
presented in Table~\ref{table:fit},  while the constraints on the radius
weaken:  $R=8.8^{+4.2}_{-0.7}$  (90 per cent confidence  limits). 
An $F$-test gives $F=3.5$ which corresponds to a 7 per cent chance 
that the decrease in $\chi^2$ was random (i.e. the improvement 
is not highly statistically significant).

We show one of the fits to the pulse  profiles  with the fixed $M$, $i$,
and $R$ in  Fig.~\ref{fig:all}a.  The  corresponding  intrinsic  angular
distribution  of the radiation  flux in the two spectral  components  is
shown in  Fig.~\ref{fig:all}b.  One sees a dramatic  difference  between
the  dependences  of the  black  body and  Comptonized  components.  The
observed  energy  dependence of the pulse profiles and the phase lags at
the pulsar frequency between  different  energies  (Fig.~\ref{fig:all}c)
are reproduced  reasonably well.  A natural  consequence of the model is
that the phase lags change  significantly at smaller  energies  together
with the black body  contribution  to the total flux, while they reach a
constant  value at about  6--8 keV where the  black  body  flux  becomes
negligible (see also Gierli\'nski et al. 2002).  
Since in reality the angular  dependence of  Comptonization
radiation  {\it is} a  function  of  energy,  one would  expect a weaker
dependence of lags on energy even above 10 keV.

\section{Discussion}
 \label{sec:disc}

\subsection{Constraints from the oscillation amplitude and the Doppler factor}
\label{sec:constr} 
\begin{figure*}
\centerline{\epsfig{file=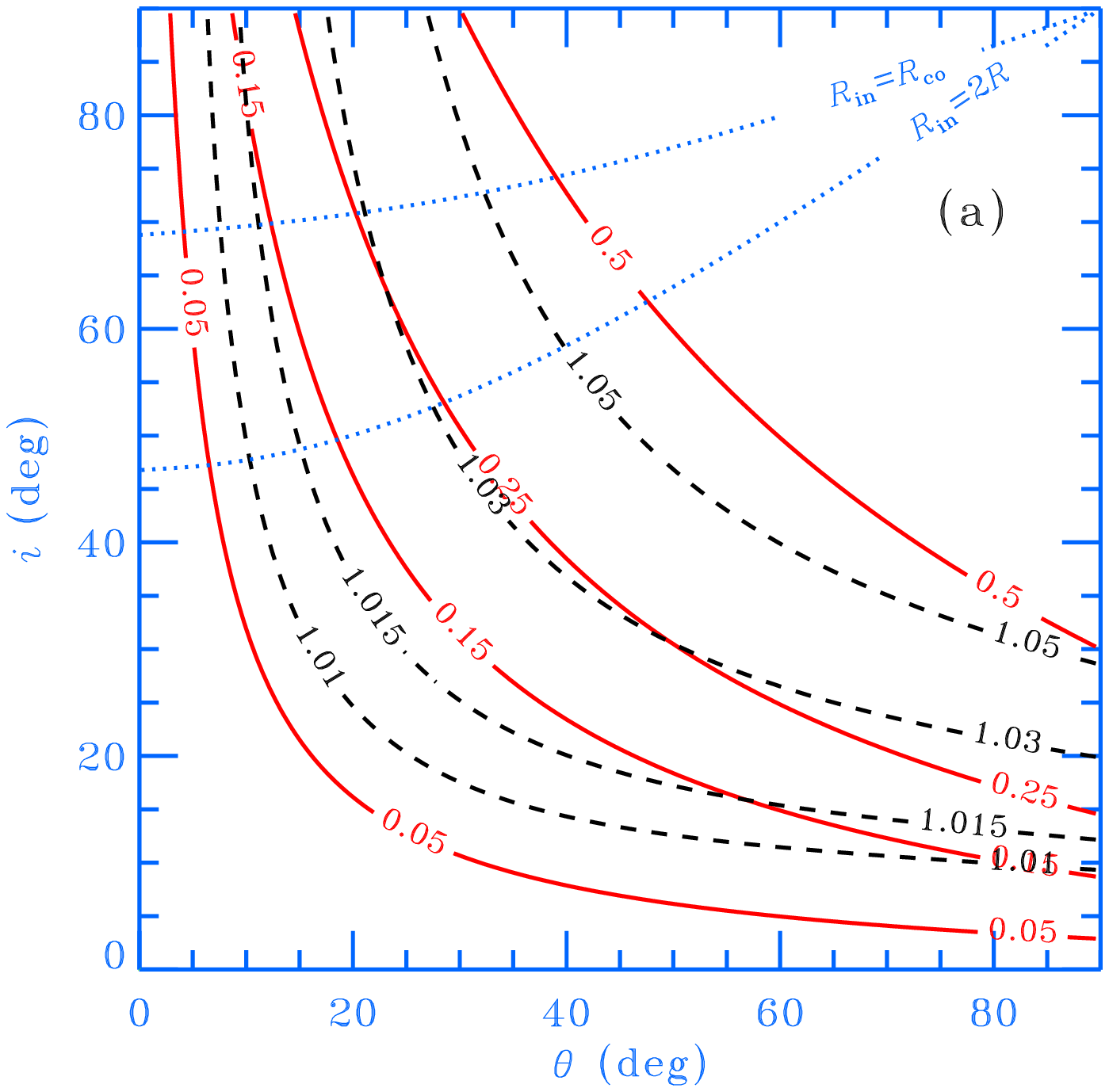,width=7.0cm}\hspace{1cm}
\epsfig{file=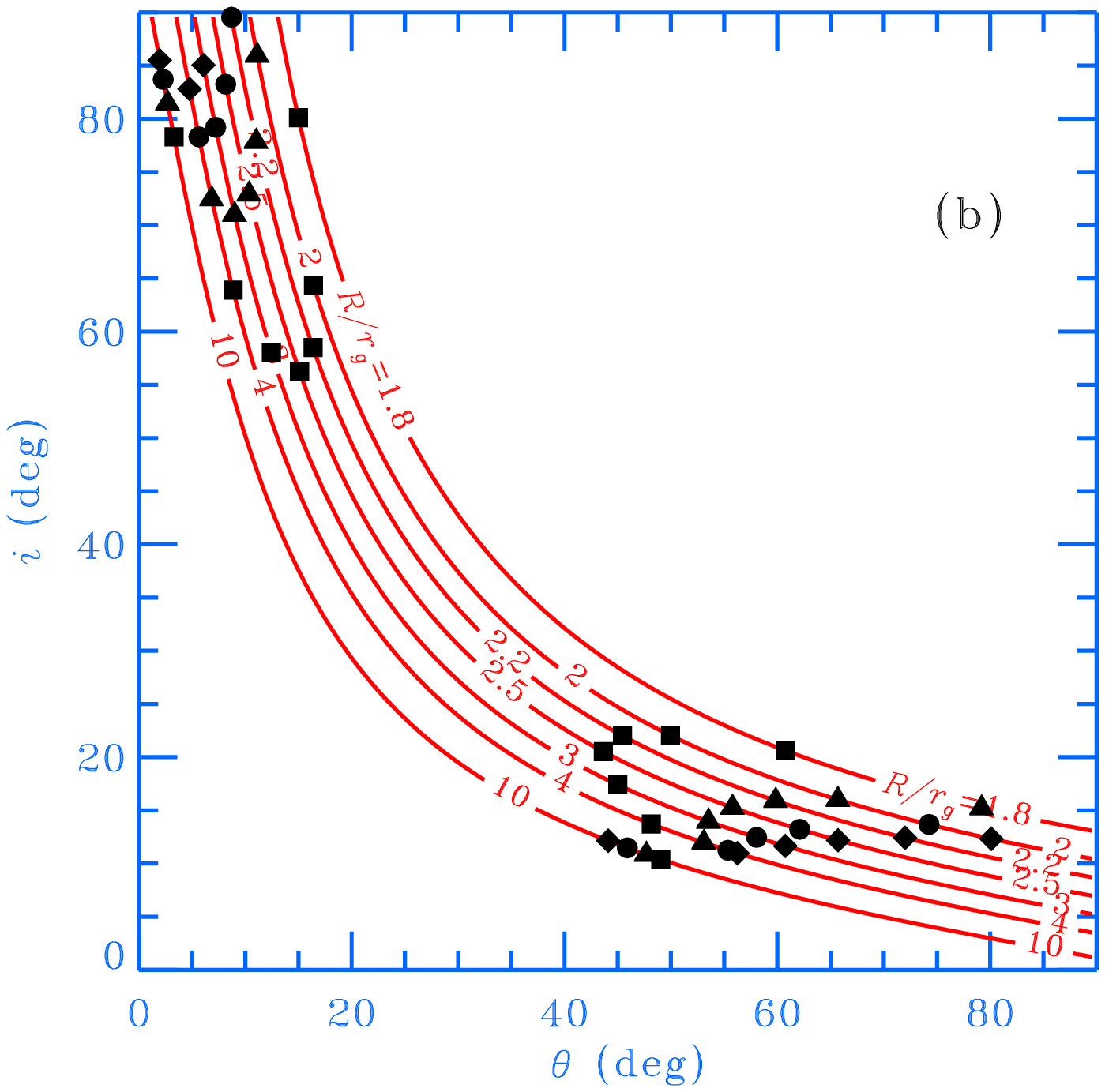,width=7.0cm}}
\caption{(a) 
Contours of the constant variability amplitude $A$ given 
by equation (\ref{eq:ampl}) (solid curves)
and the contours of constant maximum Doppler factor $\Dop_{\max}$ 
(dashed curves) for a black body spot 
($\nu=401$ Hz, $M=1.4\msun$ and $R=2\rg=8.4$ km).
The dotted curves define the $i-\theta$ relation where the photon trajectories 
from the antipodal spot (that without the disc would reach the observer) 
cross the disc at the minimum distance of $2R$ and the corotation
radius $R_{\rm co}$.
These curves are computed using approximate photon trajectories from Beloborodov (2002).
The inner disc radius smaller than $2R$ (or $R_{\rm co}$) would
block the antipodal spot from the observer in the area
above the corresponding curve. \protect\\
(b) Contours of the constant variability amplitude $A=0.18$ (solid curves) 
computed using equation (\ref{eq:ampl}) 
for different stellar radii measured in $\rg$ units  
and the interception points of these curves with the contours of 
constant $\Dop_{\max}=1.015$ for different stellar masses: 
$M=1.0\msun$ (squares), 
$1.2\msun$ (triangles), $1.4\msun$ (circles), and $1.6\msun$ (diamonds).
\label{fig:dopamrin}}
\end{figure*}

Let us consider first for simplicity  the black body emission from a single
spot. In that case,  
the  Doppler  effect  does not  change
significantly  the light  curve  (compare  dashed  and solid  curve in
Fig.~\ref{fig:relateff}a)  and thus it is mainly  characterized by the
amplitude $A$ that can be estimated for a slowly  rotating  star
using equation (\ref{eq:ampl}).  For    a
given  $R/\rg$, the  amplitude in its turn depends on $i$ and $\theta$
only.  Thus, at the  $i-\theta$  plane we can  determine  a curve that
would  correspond  to a given  amplitude  $A$  (see  solid  curves  in
Fig.~\ref{fig:dopamrin}a).  One notes that this  curve is  symmetric
around  the $i=\theta$  line  since   one can
exchange $i$ and $\theta$ in the  expression  for  $A$ 
(see eqs~[\ref{eq:ampl}--\ref{eq:uq}])
without  affecting the result.  If the black
body  emission is more beamed  towards the spot  normal because of a non-zero
$\tau$,  this curve will be somewhat  shifted  down and left since $A$
then  increases.  For example,  the black body  variability  amplitude
observed in \source\ is $\sim 25$ per cent  (Gierli\'nski  et al.  2002) and,
correcting  for $\tau$, we can determine  the region at
the $i-\theta$ plane where it can be reproduced for a given $R/\rg$.

The  situation is  different  when we consider a less beamed  emission
pattern $I(\alpha)\propto 1+a\cos\alpha$ with $a<0$ 
(i.e.  in our case  radiation  above $\sim 8$ keV).
The  oscillation  amplitude is then  strongly  affected by the Doppler
effect (compare  dashed and solid curve in  Fig.~\ref{fig:relateff}b).
The  maximum of the  Doppler  factor and the spot  projected  area are
shifted  in phase by  $\sim$  0.25 of the  period  and the peak of the
emission  is then  shifted  towards  the phase  where  $\Dop$  has the
maximum.  However, the  amplitude  and the shape of the profile  carry
the information needed to determine $a$ and the maximum Doppler factor
$\Dop_{\max}$.  Again at the  $i-\theta$  plane we can  determine  the
curve   of   constant    $\Dop_{\max}$    (see   dashed    curves   in
Fig.~\ref{fig:dopamrin}a).  Note, that since the Doppler factor (to the
first order in $\beta$)  depends on the  product  $\sin i  \sin\theta$
(see  eqs~[\ref{eq:dop}--\ref{eq:cosxi}]), but not on $i$ and $\theta$
individually,  this curve is almost  symmetric  around the  $i=\theta$
line.

A very  important  point here is that the curves of  constant  $A$ and
$\Dop_{\max}$ do not coincide (see  Fig.~\ref{fig:dopamrin}a).  At the
$i-\theta$  plane thus there are at most two points  (for a given star
mass  $M$  and  radius  $R$)  with  the  given  values  of  these  two
parameters:  points of the crossing of the curves of constant  $A$ and
$\Dop_{\max}$.  This leaves some  ambiguity  regarding the exchange of
$i$ and $\theta$.  It is possible,  however, that the number of crossing
points is one or even zero  which  would  then mean  that  there is no
physical  solution  with a given  Doppler  factor and the  variability
amplitude within the considered model.

Let us now apply our simple  arguments  to the data on  \source. The
observed light curve implies a maximum  Doppler factor about 1.015 and
the  variability  amplitude  $A$  for  a  black  body  spot  given  by
equation~(\ref{eq:ampl})  of $\sim  0.18$  (actually  observed  higher
value  results from a sharper  emission  pattern with  $\tau\sim0.1$, 
see eq.~[\ref{eq:bb}]).
The  contours of constant  $A$ do not depend on the  stellar  mass but
only on the radius  measured in $\rg$ (see  Fig.~\ref{fig:dopamrin}b).
We can now find a point at the  $i-\theta$  plane  that  satisfy  both
constraints.  The  curves  of  the  constant  $A$  and   $\Dop_{\max}$
intercept in two points:  at a large inclination and small $\theta$ as
well as at a small inclination $i\lesssim 20\degr$ and large $\theta$.

The second  solution is much less  probable.  First, we do not see any
emission from the secondary antipodal spot.  This implies that 
either it is blocked by the star itself
(which in turn  requires  the stellar radius  to be $\gtrsim 4\rg$ 
for $i\sim 15\degr$ and $\theta\sim 60\degr$),
or it is blocked  by the  accretion  disc.  
In the later  case  we  have  additional
constraints.  At  inclinations  smaller  than  $\sim50\degr$  an upper
limit on the  inner  disc  radius  (i.e.  the  magnetospheric  radius)
required to block the  secondary  spot is $R_{\rm  in}\lesssim  2R$ (see
lower  dotted  curve in  Fig.~\ref{fig:dopamrin}a).  With  such a small
inner radius, the expected amplitude of Compton  reflection  component
should be probably larger than the observed  $\Omega/2\pi\sim  0.1$ (unless the
inner disc is highly  ionised) and the shock should appear at uncomfortably
large  angle from the rotational pole if the magnetic field is a central dipole.
On the  other  hand, at inclinations $i\gtrsim 70\degr$ 
the antipodal spot is blocked by the disc if the 
magnetospheric radius is smaller than the corotation radius 
$R_{\rm co}=31 (M/1.4\msun)^{1/3}(P/2.5 {\rm ms})^{2/3}$ km which 
is also required for accretion to take place. 
Second, small $i$ is inconsistent with the X-ray 
(Chakrabarty \& Morgan 1998) and optical 
(Giles, Hill \& Greenhill 1999; Homer et al. 2001) 
modulations at the binary period, and finally, it is
also inconsistent with the lower limit $i> 28\degr$ obtained by Wang et
al.  (2001) based on the modelling of the optical/IR emission with the
X-ray irradiated disc.  We conclude that large inclinations $i>50\degr$
and small $\theta\sim  5\degr$--$20\degr$ are preferred to small $i$ and
large  $\theta$.

For small stellar  masses  $M=1.0\msun$,  the  interception  point for
given $A$ and  $\Dop_{\max}$  fixes the possible  inclination  $i$ between
$55\degr$ and $80\degr$  depending  on the assumed  radius.  The minimum
possible radius still satisfying  constraints on the inclination angle
$i<82\degr$  is  $R=1.8\rg$.  For higher  masses, the minimum  possible
radius  (in  $\rg$   units)   increases,   reaching   $R=2.5\rg$   for
$M=1.4\msun$,  and for $M=1.6\msun$  there is no solution  whatsoever.
Thus it is quite  natural that the $\chi^2$ is worse for the high mass
star (see  Table~\ref{table:fit}).  This simple analysis also explains
why the higher is the stellar mass, the higher the inclination  should
be in order to satisfy constraints on both $A$ and $\Dop_{\max}$.

\begin{figure}
\centerline{\epsfig{file=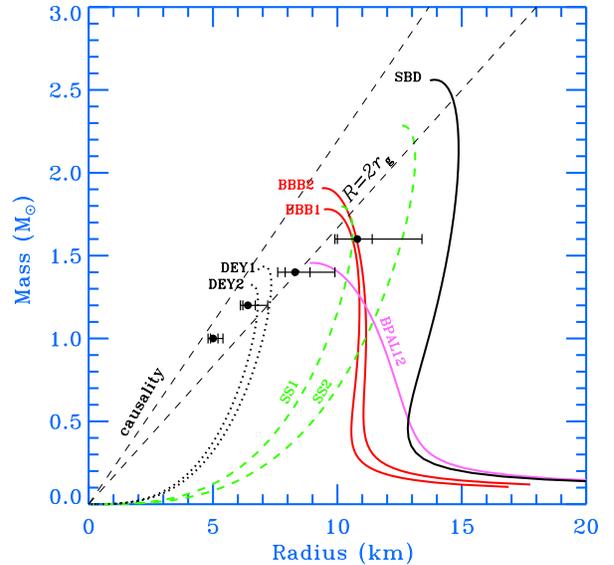,width=7.5cm}}
\caption{
Constraints  on the radius of the compact star  depending on the assumed
mass are shown by  circles  with  double  error  bars (for 90 and 99 per cent
confidence intervals) for the fits with model 2.  For comparison we show
also the  relations  between  the  compact  star mass and its radius for
several  equations  of state of neutron  stars and  strange  stars.  The
labels for neutron star  equations  of state are BBB1 and BBB2 -- Baldo,
Bombaci \& Burgio (1997), BPAL12 -- Prakash et al.  (1997), SBD -- Sahu,
Basu \& Datta (1993), while for the strange stars SS1 and SS2 correspond
to the MIT bag model (see  Gondek-Rosi\'nska,  Klu\'zniak \& Stergioulas
2003 for  details) and DEY1 and DEY2 to the  equation of state of Dey et
al.  (1998) computed as in Gondek-Rosi\'nska et al.  (2000).  
All realistic equations of state lie to the right of the 
causality  line.  
The line $R=2\rg$ is shown to guide the eye.
\label{fig:eos}}
\end{figure}

\subsection{Stellar masses and radii}
\label{sec:massrad}

Our results (see Table~\ref{table:fit} and Fig.~\ref{fig:eos}) 
put constraints on mass 
of the compact object. For $M=1\msun$ the obtained  stellar  radius of 
$R\sim 5$ km is too  small to be consistent with any existing 
equations of state for neutron or strange stars. For masses higher 
than $1.6\msun$, the fits are much worse than for smaller masses. 
Thus we conclude  that the stellar mass is bound  between 1.2
and 1.6 solar masses. 

We can also put constraints on the equations of state of compact stars 
(which of course could be much stronger if the mass of compact object were known).
Thus, if $M=1.2\msun$ the only possible solution is a strange star
(Chakrabarty  1991; Li et al.  1999; Ray et al.  2000;
Gondek-Rosi\'nska  et al.  2000)
with the equation of state derived by Dey et al. (1998).
For the mass about $1.4\msun$,
almost all known neutron star equations of state
predict larger radii than our best-fitting value. The exception
are the models similar to BPAL12 from Prakash et al. (1997) and
the models with a kaon condensate (Glendenning \& Schaffner-Bielich 1999;
see Lattimer \& Prakash 2001 for more details).
However, the maximum mass for these models is $\sim 1.5\msun$ which
is smaller than the best estimate of the neutron star mass 
in Vela X-1 of $1.86\pm0.16\msun$  (Barziv et al. 2001).
Thus we are inclined to conclude that 
our constraints together with the minimum neutron star mass in 
Vela X-1 put a lower limit on to the neutron  star  mass in 
\source\ of about $1.5\msun$.
Strange stars (MIT bag models)
with maximum masses in the range of $1.5$--$1.7\msun$
would also fit into our constraints.
At larger mass of $M=1.6\msun$ a strange star is still a
possible solution as well as a neutron star described by 
equations of state similar to that of Baldo et al. (1997), 
while the most stiff equations of state, such as for example 
that by Sahu et al. (1993), can be ruled out (see Fig.~\ref{fig:eos}).

Assuming alignment  of the orbital angular momentum and that of the 
compact object, we can estimate the mass of the companion star $M_c$ 
from our lower limits on the inclination of the rotational axis (see Table~\ref{table:fit}).
Using the measured mass function 
$f_x=(M_c \sin\ i)^3/(M+M_c)^2=3.78\times 10^{-5}\msun$
(Chakrabarty \& Morgan  1998) we find that 
$M_c=0.041\pm0.002, 0.0445\pm0.0015, 0.0478\pm0.0005\msun$
for $M=1.2, 1.4, 1.6\msun$, respectively. 
The companion radius is about $0.13R_{\odot}$ (Bildsten 
\& Chakrabarty 2001). 

The mass measurements of radio pulsars in binary systems and their 
neutron star companions  revealed their very narrow distribution with the
mean $M\sim1.35\msun$ and a very small dispersion (Thorsett \& Chakrabarty 1999).
No evidence for extensive mass accretion was found. 
In order to spin up a star with the moment of inertia $\sim10^{45}$ g cm$^2$ 
to the frequency of 400 Hz one needs to accrete just about
$\Delta M\approx 0.1\msun$. Thus if the initial mass is within 
the same distribution one does not expect the mass of the 
compact object to exceed $1.5$--$1.6\msun$. 
For a mass of $1.4\msun$, our best-fitting 
radius (with model 2) is $R=8.3^{+0.6}_{-0.4}$ km (90 per cent 
confidence interval).
Such a radius is consistent with the 
upper limit $R<7.5 (M/\msun)^{1/3}$ km 
obtained from the fact 
that pulsations are observed at fluxes differing by a
factor of 100 (Burderi \& King 1998; Li et al. 1999). 
Here 
one assumes that the magnetic  and the ram pressure from the accreting
material are equal at the magnetospheric radius which is larger than 
the stellar radius $R$ and smaller than the corotation radius $R_{\rm co}$. 
For $M=1.5\msun$, our best fit $R=9.5\pm0.5$ km is already inconsistent with 
the given upper limit.  
This constraint, however, is more relaxed if the 
magnetic field does not have a dipole structure 
and/or the accretion disc is not gas-pressure dominated 
(Psaltis \& Chakrabarty 1999).

\subsection{Emission mechanism}

The  angular dependence of the  hard radiation
that is needed to produce the observed pulse
profiles above 8 keV is consistent with that expected from
the electron scattering dominated slab  with Thomson
optical depth $\taut\sim 0.3$--$1$.
It would be natural to assume that the emission originates in
a plane-parallel shock at the surface of the compact star (Basko \& Sunyaev 1976).
The observed spectral slope $\Gamma\sim2$ (the energy spectrum $E^{-1}$)
is expected from bulk motion Comptonization in a
radiation dominated strong shock (Blandford \& Payne 1981). 
Most of the papers on this subject neglect generation of new 
soft photons by reprocessing the hard radiation at the neutron star 
surface which would significantly soften the emergent spectra.
Thus, in order to reproduce $\Gamma\sim2$ slope, 
thermal Comptonization should play an important role 
(Lyubarskii \& Sunyaev 1982). As discussed in Gilfanov et al. (1998),
there should exist a mechanism that adjust the spectrum so that it
does not change much when the accretion rate varies by a factor of 100.
Reprocessing of the hard photons produced in a shocked region into 
soft ones that cool the shock could be such a mechanism.
This  two-phase structure in the energy balance is known to stabilize
the spectrum  at the observed values when the optical depth is of the order of unity
(Haardt \&  Maraschi  1993;  Stern  et  al.  1995).
The  electron  temperature adjusts itself to the variations
of the  optical  depth  (and  luminosity)  so that  the  product  of the
electron temperature and optical depth is approximately constant.

Variation of the accretion rate that is the main candidate for the
aperiodic variability observed in \source\ (Wijnands \& van der
Klis 1998b) thus can cause the changes in the optical depth,
but not in the spectral shape.  However, the relative amplitude of the black
body and Comptonized components can vary.  Indeed, at small luminosities
the black body seems to the more prominent (Gilfanov et al. 1998) which
would correspond to a smaller optical depth of the Comptonizing layer.
A larger relative normalization of the black body component 
should result also in a higher variability amplitude 
which in fact was increasing from 4 to 7 per cent (rms amplitude) 
over the duration of the 1998 April outburst (Cui et al. 1998).

There could be additional effects related to the electron-positron pair production 
(by photon-photon interactions) 
which becomes important when the compactness parameter
$l=L\sigma_T/(\rspotc\ m_ec^3)$ is larger than unity.
For the luminosity of $L\sim 3\times 10^{36}$ erg s$^{-1}$ (which 
is a characteristic luminosity during the  April 1998 outburst 
of \source, see Gierli\'nski et al. 2002) and the 
emission region size of $\rspotc=2$ km, the compactness
is about $400$ and  it could be much larger if the region has a
smaller height. Thus, the data are consistent with the presence 
of pairs which could be responsible for 
producing the optical depth in the range $\taut\sim 0.3-1$ 
depending on the compactness and for fixing the electron temperature 
in the range $50-100$ keV working as a thermostat 
(Stern et al. 1995; Poutanen \& Svensson 1996; 
Malzac, Beloborodov \& Poutanen 2001). 
If the hard radiation is slightly beamed towards the neutron 
star surface due to the bulk motion, the amount of soft 
photons produced would be larger than from 
an isotropic source and there will be no problem in
reproducing the observed spectrum that satisfies the energy 
balance (cf.  Gierli\'nski et al. 2002).

The radiation escaping from the shock can be affected by cold plasma 
in the accretion column. Assuming the area of the emission region of
$\pi \rspotc^2\sim12$ km$^2$  and the accretion rate
$\dot{M}=2\times 10^{16}$ g s$^{-1}$
(needed to produce the observed luminosity at efficiency of 0.15)
one can estimate from the continuity equation
the characteristic optical depth of the plasma over the distance equal to 
the spot radius $\tau=n_e \sigma_T \rspotc \approx 0.8$.
(At higher distances from the star, the magnetic field lines 
diverge and photons can escape freely.)
This means that the flow can affect the radiation
escaping along the normal to the spot and the radiation can
be blocked from the observer at certain phases.
Since the area emitting the black body radiation
is larger than  the cross-section of the accretion column,
one would expect that the effect of shadowing is smaller for
the black body.  Possibly, this effect is
observed in the light curve where the hard radiation
has a break at $\phi=0$ (see Fig.~\ref{fig:all}a).
We may also see this effect from the best-fitting angular 
distribution (see Fig.~\ref{fig:all}b) which shows a drop along 
the normal at 
$\mu=\cos\alpha'\sim 1$.

\subsection{Origin of soft lags} 

Our results confirm the proposal by  Gierli\'nski et al. (2002)
that the main physical reason behind  the soft lags is a two-component 
nature of the spectrum. A natural consequence of this model
is a break in the phase lag at the energy where 
the black body component vanishes, i.e. around 8 keV 
(see Fig.~\ref{fig:all}c). 
We showed here that the black body and the Comptonized 
components having different 
angular distribution generate light curves which are affected 
by the Doppler effect in a different way. 
The Doppler shift is an important ingredient but not the major 
cause of the soft lags. In our model, the Doppler effect comes
into play only through its influence on the variability amplitude 
of the {\it bolometric} flux. 

On the other hand,
Ford (2000) and Weinberg et al. (2001)  proposed a model
where the lags in \source\ are produced  by the Doppler effect 
alone through its influence on the {\it monochromatic} flux. 
In both  papers  a black  body emission from the spot is  assumed. 
When the spot is moving towards the observer, 
the Doppler effect shift the spectrum towards higher energies, while 
motions away from the observer softens the spectrum and 
generate a deficit at high energies. Thus, higher energy photons,
in the tail of the black body distribution, 
arrive {\it only} at phases when motion is predominantly towards
the observer, while the flux of photons at the peak of the black body distribution
has a maximum together with the projected area, i.e. quarter of the period later. 
This results in hard leads (or soft lags) as well.
As discussed in Gierli\'nski et al. (2002), there are 
a number of problems with this proposal. 
First, the photons in the black body tail contribute 
very little to the total observed spectrum above 8 keV which 
is dominated by the hard Comptonized component. Thus, any soft lags
in the black body flux would be completely invisible in the total flux.
Second, a break in the phase lag energy dependence 
does not have any physical explanation in that model.

Even stronger Doppler boosting is expected when photons from the spot 
are scattered off a much more rapidly rotating inner disc 
(Sazonov \& Sunyaev 2001). The scattered radiation will be 
leading or lagging the incident spot radiation depending on whether 
the disc is co-rotating or counter-rotating.
For a black body radiation from the spot, this would translate into 
soft lags or leads, respectively. However, since the spectrum 
from the spot is closer to a power law, no phase lags should appear
(Chen \& Shaham 1989). A different angular distribution of the
black body and Comptonized components as well as 
a non-coherent nature of scattering can affect this result. 
It could be an interesting problem for future studies. 
We believe that this model is not directly related to \source,
since the spot is probably situated close to the rotational pole
and the fraction of the spot photons scattered in the disc 
is negligibly small.

Cui et al.  (1998) interpreted soft lags as due to
Compton down-scattering   of intrinsic hard photons
with energies above 10 keV 
in a cold electron cloud of Thomson optical depth $\taut\sim 10$.  The
softer  photons  (scattered  more times)  escape  from the medium  later
producing  soft lags.  As the authors pointed out, the hard photons
should escape to the observer directly through some kind of a hole
in the cloud since otherwise a spectral cutoff  should appear at
$\sim 511/\taut^2\sim 5$ keV (e.g. Sunyaev \&  Titarchuk 1980;  
Lightman, Lamb \& Rybicki 1981). This model, however, 
predicts a  significantly reduced variability in the softer  energy band
(Chang \& Kylafis 1983; Brainerd \& Lamb 1987) contrary  to  what is observed.

\section{Summary}

\label{sec:summary}

The two main  spectral  components,  black  body  and  Comptonized,  are
clearly identified in the time-averaged  spectrum of \source.  The pulse
profiles show significant  energy  dependence and the softer photons lag
the  harder  ones.  Such a  behaviour  can be  reproduced  if  the  hard
Comptonized   emission  peaks  earlier  than  the  black  body  emission
(Gierli\'nski  et al.  2002).  The black body component shows about 25 per cent
(peak-to-peak)
variability  amplitude  and the Doppler  effect  (producing a $\sim$ 7 per cent
variability)    does    not    change    its    pulse    profile    much
(Fig.~\ref{fig:relateff}).  The  Comptonized  emission from an optically
thin slab with a broader  (``fan''--like)  angular  distribution  has an
intrinsically  smaller  variability  amplitude.  The Doppler effect then
shifts significantly the peak of the pulse producing soft lags.

By modelling the observed pulse  profiles, we obtained an upper limit
on the mass of the compact object  $M\lesssim1.6\msun$. We also put the lower limit
of $M\gtrsim1.2\msun$ on the stellar mass since for smaller masses the obtained
radii are too small to be consistent  with any
published equation of state for neutron or strange stars. This also
constrains the inclination of the system to be larger than  $65\degr$.
For the  masses
$1.4\msun<M<1.6\msun$, the best-fitting stellar radii $R=8$--$11$ km are
consistent  with those given by the neutron  star  equations  of states
as well as some strange star models,
while for $M\sim 1.2\msun$ with that given by the equations of state for
strange stars only.

\section*{Acknowledgments}

We thank Andrei  Beloborodov, Tomasz Bulik, Marat Gilfanov, George Pavlov and Boris Stern
for useful  discussions.  We are grateful to Marat  Gilfanov for
providing the pulse profiles of \source\ for comparison  and to Dorota
Gondek-Rosi\'nska  for the  data on the  equations  of  state  for the
neutron and strange stars.  This research has been supported by the Academy
of Finland and the Jenny and Antti Wihuri Foundation. JP is grateful
to NORDITA (Copenhagen) and to  Didier
Barret at Centre d'Etude Spatiale des Rayonnements (Toulouse)
 for the hospitality during his visits.

\label{lastpage}
\end{document}